\documentclass[%
 reprint,
 nofootinbib,
 amsmath,amssymb,
 superscriptaddress,
 aps,
]{revtex4-2}

\usepackage{graphicx}
\usepackage{dcolumn}
\usepackage{bm}
\usepackage{subcaption}
\usepackage{lipsum}
\usepackage{tabularx}
\usepackage{url}
\usepackage{array, blkarray, makecell}
\usepackage{tabularx}
\setcellgapes{6pt}

\preprint{APS/123-QED}

\begin{document}

\title{Measurement of Ionization Produced by 254~eV\textsubscript{nr} Nuclear Recoils in Germanium}

\author{A.R.L. Kavner}
\email{akavner@umich.edu}
\affiliation{Applied Physics Program, University of Michigan, Ann Arbor, MI, USA}

\author{I. Jovanovic}
\affiliation{Applied Physics Program, University of Michigan, Ann Arbor, MI, USA}
\affiliation{Department of Nuclear Engineering and Radiological Sciences, University of Michigan, Ann Arbor, MI, USA}

\date{\today}

\begin{abstract}
Ionization produced by low-energy nuclear recoils is among the primary direct signatures of dark matter interactions. Despite the urgency of dark matter detection and the recent measurements of coherent elastic neutrino-nucleus scattering, detector response to nuclear recoils is not well characterized in the keV\textsubscript{nr} and sub-keV\textsubscript{nr} regime across a variety of materials. We have re-performed a measurement of the ionization produced by monoenergetic 254~eV\textsubscript{nr} nuclear recoils in Ge with improved digital electronics and additional systematic studies. Our results indicate an ionization yield of $64\pm8$~eV\textsubscript{ee} corresponding to a quenching factor of $25\pm3$\%, greater than the 14\% predicted by the Lindhard model. This ionization enhancement could greatly improve the sensitivity of high-purity Ge detectors in dark matter detection and measurement of neutrinos via coherent scattering.
\end{abstract}

\maketitle

\section{\label{sec:Intro}Introduction}

    \subsection{Dark Matter and Neutrino Detection with High Purity Germanium}

    Direct detection of dark matter is among the highest priorities in experimental cosmology and particle physics~\cite{Kolb2018,Feng2023,Snowmass2021}. Measurement of dark matter requires detectors with sensitivity to low-energy nuclear recoils $\lesssim$1~keV\textsubscript{nr} as well as an understanding of the detector response~\cite{Xu2023}. The necessity for accurate detector response models to nuclear recoils has increased with the recent direct detection of coherent elastic neutrino-nucleus scattering (CE$\nu$NS)~\cite{Akimov_2017,CsI2019}. It has been shown that the choice of quenching factor model greatly affects the degree to which experimental data agrees with Standard Model predictions~\cite{CsI2019}.

    The application of high-purity germanium (HPGe) detectors for dark matter searches, CE$\nu$NS detection, and other rare-event physics experiments has increased in popularity~\cite{Fourches19}. This trend is due in part to the commercial availability of multi-kilogram HPGe detectors with suitably low noise and low backgrounds~\cite{Barbeau_2007,CoGeNT2011,Cogent_2013,C4,ChicagoReactor_2021,CONUS2021,Scovell_2024}. Multiple collaborations have or plan to deploy HPGe detectors for experiments aimed at detecting dark matter or for CE$\nu$NS based neutrino experiments ~\cite{Barbeau_2007,Cogent_2013,C4,Akimov_2017,Scholberg_2020,CONUS2021,ChicagoReactor_2021,ChicagoReactor_2022,CEDEX10,TEXONO,ackermann2024,PhysRevD.106.L051101}. The CDEX collaboration is investigating the dark matter sensitivity of a 50~kg array of HPGe detectors~\cite{CDEX50}. Given these advancements, a dark matter search with greater than 100~kg of detection material could be considered in the near future. 

    \subsection{Lindhard Model}

    Detector response to nuclear recoils is conventionally expressed in terms of the quenching factor~\cite{Snowmass2021,Xu2023}. This is defined as the ratio of the measured ionization (or scintillation) energy, denoted electron equivalent with subscript ``ee'' as a fraction of the kinetic energy of the recoiling nucleus, denoted with subscript ``nr''. In crystals such as Ge, the relationship between detectable electron equivalent energy (E\textsubscript{ee}) and nuclear recoil energy (E\textsubscript{nr}) is given by the Lindhard model~\cite{Lindhard}:
    \begin{align}
        \centering
        E_{\text{ee}} &= \frac{\kappa g(\epsilon)}{1 + \kappa g(\epsilon)}\cdot E_{\text{nr}}\label{eq:Lindhard1}\\
        g(\epsilon) &= 3\epsilon^{0.15} + 0.7\epsilon^{0.6} + \epsilon \label{eq:Lindhard2}\\
        \epsilon &= 11.5Z^{-7/3}E_{\text{nr}} \label{eq:Lindhard3}
    \end{align}
    The term $\kappa$ is theoretically predicted to be 0.157 for germanium~\cite{Lindhard}. \textit{Z} is the atomic number, 32 in the case of Ge. 
    
    Experimental data agrees well with $\kappa=0.157$ above $\sim$10~keV\textsubscript{nr}~\cite{BAUDIS1998348,Jones1965,Jones1967,MESSOUS1995}. Below $\sim$10~keV\textsubscript{nr}, numerous experiments have reported results with a best-fit value of $\kappa>0.157$ or significant disagreement with the predictions of Lindhard theory~\cite{Bonhomme2022,Scholz2016,MESSOUS1995,GeQF2021,TEXONO,Xu2023}. Other works have reported lower ionization than expected by Lindhard below 10~keV\textsubscript{nr}, though these experiments were performed at millikelvin temperatures~\cite{CDMS2011,CDMS2022,CDMS2010QF}.

    \subsection{Low Energy Discrepancy}

    A recent multi-pronged study performed by the University of Chicago (UChicago) extensively studied the HPGe detector response to low-energy nuclear recoils and reported a greater quenching factor than predicted by the Lindhard model for recoils with energy $\lesssim1.5$~keV\textsubscript{nr}~\cite{GeQF2021,Lindhard}.
    Of particular interest is the measurement of the ionization produced by monoenergetic 254~eV\textsubscript{nr} $^{73}$Ge nuclei. The UChicago study reported a 44\% greater ionization yield\footnote[1]{The value of 44~\% is calculated using gamma-ray energy of 68.752~keV and values of 68.811 and 68.793~keV for the gamma-ray plus ionization produced by the nuclear recoil~\cite{GeQF2021}. See Table~\ref{tab:results}.} than an earlier Brookhaven National Laboratory based experiment~\cite{GeQF2021,Jones1975}. The Brookhaven result is consistent with the ionization yield predicted by the Lindhard model~\cite{Jones1975,Lindhard} while the UChicago result is in tension~\cite{GeQF2021}.
    %

    To address this discrepancy, we have re-performed the measurement in the same setting as the UChicago study. We have improved the experimental setup and procedure by utilizing modern digital electronics and saving the raw detector waveforms. Additional systematic studies were performed including a gamma tagging measurement, an investigation into the effects of signal processing, and a study of the effect of interaction location within the detector volume. Our results corroborate the UChicago study~\cite{GeQF2021}, in disagreement with the earlier Brookhaven experiment~\cite{Jones1975}. 

    %
    
\section{Monoenergetic 254~eV\textsubscript{nr} Nuclear Recoils}

    \subsection{Level Structure of $^{73m}$Ge}
\begin{figure}[h]
        \includegraphics[width=.75\linewidth]{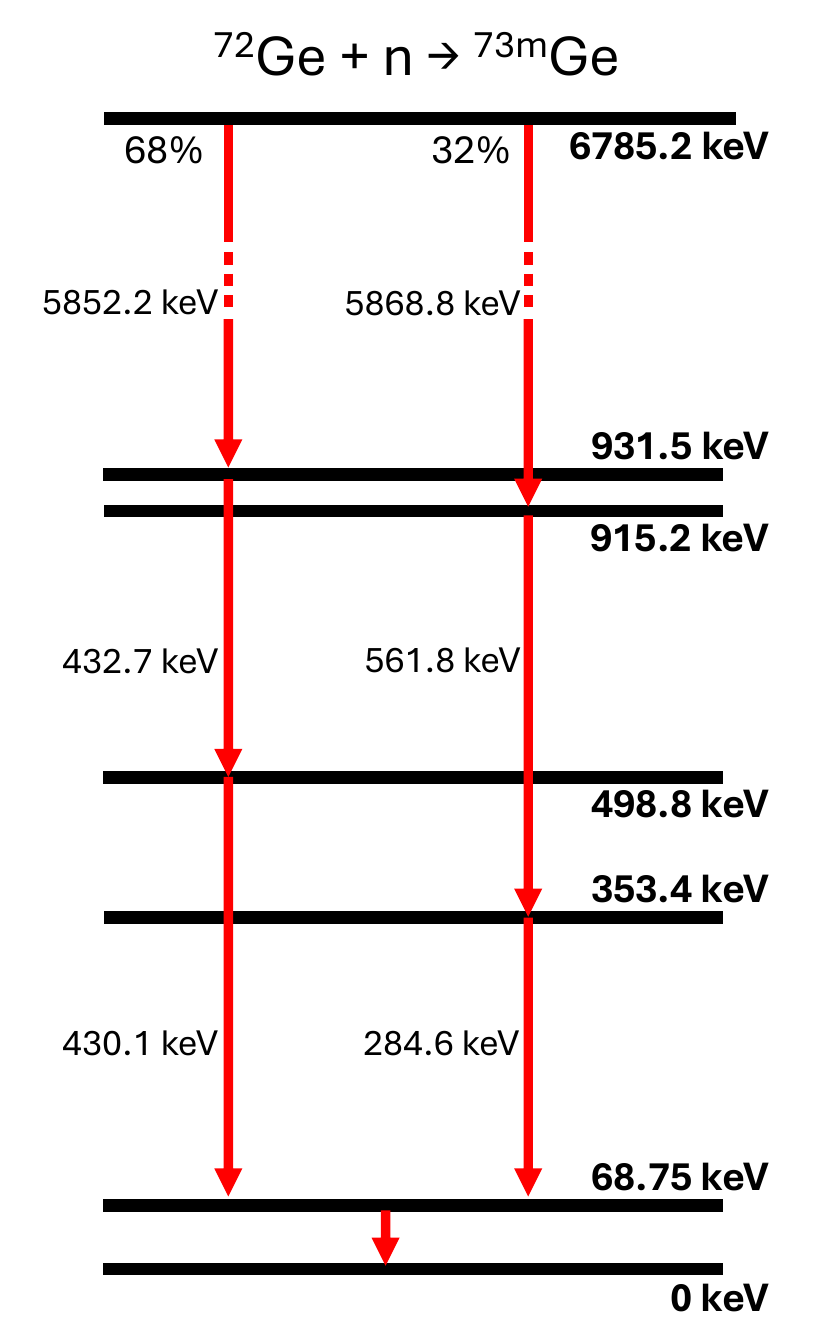}
        \caption{\label{fig:GeLevels} De-excitation path of $^{73m}$Ge that feeds the 68.75~keV level. State energies are listed on the right while gamma-ray energies are listed left of the red arrows designating the de-excitation path.}
    \end{figure}

    Monoenergetic 254~eV\textsubscript{nr} nuclear recoils are produced via the emission of 5.9~MeV gamma rays following thermal neutron capture. Thermal neutron capture on $^{72}$Ge, comprising 27.4\% of natural Ge, populates the 6785.2~keV excited state of $^{73m}$Ge~\cite{PhysRevC.43.1086,Jones1971,Jones1975}. The decay path of interest is depicted in Figure~\ref{fig:GeLevels}, where the majority of nuclear excitation is radiated via the emissions of 5852.2~keV or 5868.8~keV gamma rays that feed into the 915.2~keV or the 931.5~keV level, respectively. Only the 915.2~keV and 931.5~keV levels feed into the subsequent 68.75~keV state, which then decays to the ground state of $^{73}$Ge~\cite{PhysRevC.43.1086,Jones1971,Jones1975,NDS,IAEANDS}. 

    The de-excitation of $^{73m}$Ge results in nuclear recoils by conservation of momentum. Emission of the 5852.2~keV and 5868.8~keV gamma rays produce 253.5~eV\textsubscript{nr} and 252.1~eV\textsubscript{nr} nuclear recoils, respectively. The other gamma rays emitted contribute negligibly ($\lesssim$1\%) to the total nuclear recoil energy. The prior study~\cite{Jones1975} calculated the intensity-weighted average recoil energy of 254.1~eV\textsubscript{nr} with a spread of 1.5~eV\textsubscript{nr}\footnote[2]{The calculation assumes rapid stopping of the $^{73}$Ge nucleus compared to the lifetimes of nuclear states.}.

    All gamma rays released in de-excitation of $^{73m}$Ge, except the lowest energy 68.75~keV gamma, have a high probability of escaping a small 2~cm$^{3}$ HPGe crystal~\footnote[3]{1.6~cm diameter $\times$ 1~cm height is the crystal geometry of the Ortec GLP-16195/10P4 detector used in our experiment.} without interaction; their attenuation lengths in Ge are all longer than a centimeter. Attenuation lengths are calculated from the XCOM (NIST) database~\cite{XCOM} and are listed in Table~\ref{tab:Levels}. 

    Simulations in MCNPX~\cite{MCNP} framework were used to model a uniform gamma-ray source emitted from a 1.6~cm~(diameter) $\times$ 1.0~cm (height) cylindrical Ge crystal, the geometry used in our experiment. The escape fraction, which we define as the fraction of gamma rays for a given emission line that do not interact with the crystal, either photoelectrically or by Compton scattering, is given in Table~\ref{tab:Levels}. The combined probability that none of the gamma rays preceding the 68.75~keV level interact with the crystal is 30.35\%, and the probability of a 68.75~keV gamma ray being photoelectrically absorbed within the crystal volume is 84.82\%. The combined 68.75~keV gamma ray and nuclear recoil energy deposition is detected with an efficiency of 25.5\%.

    %
    %
    %

    \begin{table}[h]
        \centering
        \begin{tabular}{|c|c|c|c|}
        \hline
        ~~~~~~E\textsubscript{$\gamma$}~~~~~~  & ~~~~~~E\textsubscript{nr}~~~~~~        & Attenuation &  Escape Fraction \\ 
        ~[keV]                      &  [eV\textsubscript{nr}]    & [mm]       &  [\%]          \\
        \hline
        \hline
        ~5868.8 & 253.5 &  60.4  & 91.3   \\
        ~5852.2 & 252.1 &  60.3 &   91.3   \\
        ~~561.8  &  2.3  &  24.3 &  80.7  \\
        ~~432.7  &  1.4  &  21.1 & 78.4 \\
        ~~430.1  &  1.4  &  21.0    &  78.3    \\
        ~~284.6  &  0.6  & 15.9 &   73.5   \\
        ~~68.75  &  0.0  &  1.3  &  15.2  \\
        \hline
        \end{tabular}
        \caption{Emitted gamma rays from de-excitation of $^{73m}$Ge which decay to ground state via the 68.75~keV state. Attenuation lengths are calculated from the XCOM NIST database~\cite{XCOM}. The escape fraction is defined as the fraction of gamma rays that do not interact with the crystal and is determined by Monte-Carlo simulation assuming a uniform source in a 2~cm$^{3}$ crystal used in the experiment.}
        \label{tab:Levels}
    \end{table}

    \subsection{Energy of the 68.75~keV Level}
    
    In our experiment, as in the prior UChicago and Brookhaven experiments, the ionization produced by the 254~eV\textsubscript{nr} nuclear recoil is measured summed with the energy deposited by the final 68.75~keV gamma ray. The ionization energy is calculated from the difference between this summed energy and the energy of the final state. Accurate interpretation of the quenching factor therefore requires precise knowledge of the 68.75~keV gamma ray. 

    The Brookhaven experiment reported a gamma-ray energy of 68.7535$\pm$0.0043~keV~\cite{Jones1975}. The UChicago study measured the gamma-ray energy to be 68.734~keV with an uncertainty of $\pm$20~eV~\cite{GeQF2021}. The IAEA and ENSDF database list the gamma-ray energy to be 68.752$\pm$0.007~keV for which numerous sources, including the Brookhaven study, are cited~\cite{NDS,IAEANDS,SALZMAN1972312,Forssten_1974,PhysRevC.92.054302,PhysRevLett.32.512, Heymann1969, Forssten_1976}. We adopt this ENSDF value from which we base our calculations. 

    %
    %
    \subsection{Prior Results}

    To the best of our knowledge, Refs.~\cite{Jones1975} and~\cite{GeQF2021} are the only two prior measurements of the germanium quenching factor at 254~eV\textsubscript{nr}. This was achieved by irradiating germanium detectors with strong sources of thermal neutrons at Brookhaven National Laboratory and the Ohio State University Nuclear Reactor Laboratory, respectively. 

    The Brookhaven experiment measured a gamma plus recoil combined energy of 68.7927$\pm$0.0034~keV. Subtracting the gamma-ray energy of 68.752$\pm$0.007~keV gives an ionization yield of 40.7$\pm$7.8~eV\textsubscript{ee} corresponding to a quenching factor of 16.0$\pm$3\%. This value is consistent with the Lindhard prediction of 14.3\% for $\kappa=0.157$. 

    The UChicago study measured a larger value of 68.811$\pm$0.001~keV for the gamma-ray plus nuclear recoil signal. If the same 68.752$\pm$0.007~keV is subtracted, this gives an ionization yield of 59.0$\pm$7.1~eV\textsubscript{ee} and a quenching factor of 23.2$\pm$2.8\%. The UChicago work hypothesizes the difference between their results and the Brookhaven result was caused by the 700~ns lifetime of the 68.75~keV state and the difference in choice of time constants of their analog shaping amplifiers, 8~\textmu s and 2~\textmu s for UChicago and Brookhaven respectively~\cite{GeQF2021,Jones1975}.

    It should be noted that in Ref.~\cite{GeQF2021} a gamma ray energy of 68.734~keV is subtracted yielding a quenching factor of 30.3$\pm$7.9~\%. This difference highlights the importance of accurate knowledge of the 68.75~keV gamma ray for the quenching factor measurement. 

    Regardless of the choice of gamma-ray energy, the UChicago result is significantly greater than the ionization predicted by Lindhard theory~\cite{Lindhard}. This discrepancy motivates additional studies since such an enhanced quenching factor would significantly improve the sensitivity of HPGe detectors to low-energy recoils for dark matter and reactor CE$\nu$NS detection.

\section{Experimental Methodology}

    \subsection{Neutron Source}

    The experiment was deployed at the Ohio State University Nuclear Reactor Laboratory~\cite{OSULab}. The reactor was operated at an estimated thermal neutron flux of $1-2 \times 10^6$~cm$^{-2}$ s$^{-1}$ and was chosen for the high thermal purity of the neutron beam. The thermal neutron facility has been well characterized; only 3.76 neutrons in every 1000 delivered in the thermal beam have energy above 0.4~eV~\cite{OSU2012}. Above thermal energies, the neutron spectrum is flat with a small peak above 1~MeV. The neutron energy spectrum and the spatial extent of the beam are described in Figures~4 and~7 of Ref.~\cite{OSU2012}, respectively. 
    Thermal neutron purity is critical as any momentum imparted by inelastic scattering would result in an erroneously high measured value of the quenching factor. The facility is the same one used in the UChicago study~\cite{GeQF2021}.

    \begin{figure}[h]
        \includegraphics[width=.95\linewidth]{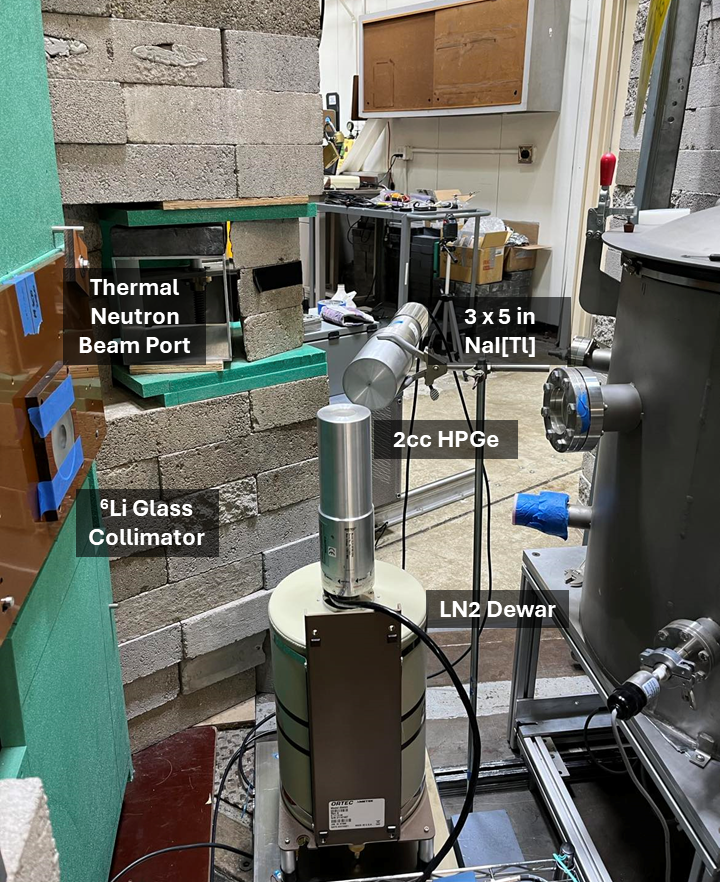}
        \caption{Labeled image of the experimental setup at the OSU nuclear reactor laboratory. The HPGe detector was placed within the collimated neutron beam while the NaI(Tl) detector was placed to the side outside the neutron beam.}\label{fig:setup}
    \end{figure}
    %
    %
    %

    \subsection{Detectors}
    
    The experimental apparatus comprised two radiation detectors: the HPGe detector, which acted as the neutron beam target, and a large external scintillation detector to tag the emitted 5.8~MeV gamma ray. The HPGe detector was an Ortec GLP-16195/10P4 detector in PopTop configuration mounted on a multi-orientation portable cryostat~\cite{Ortec,GLP}. The HPGe crystal volume was 2~cm$^{3}$, mass 10.7~g, and cooled to 77~K. The crystal is comparable in volume to the 1~cm$^{3}$ HPGe crystal utilized in the UChicago study~\cite{GeQF2021} and 1~cm$^{3}$ Ge(Li) crystal from the Brookhaven measurement~\cite{Jones1975}.

    The outward-facing ion-implanted layer of the GLP detector is only 0.3~\textmu m thick and has negligible attenuation for externally incident X rays and gamma rays~\cite{GLP,GeQF2021,CANNIZZARO2001129}. The GLP detector series is commonly used for measurement and spectroscopy of X rays and gamma rays with energies between 1 and 200~keV~\cite{CANNIZZARO2001129}, an ideal match to the energy of interest in this experiment.

    The HPGe detector was placed in the reactor thermal neutron beam as depicted in Figure~\ref{fig:setup}. A 1~cm thick $^{6}$Li glass disc with a 1~inch diameter hole was used to collimate the beam and reduce activation backgrounds. A null measurement was performed by placing a different $^{6}$Li disc with no hole, which completely blocked the beam port.

    A large 3~inch diameter, 5~inch long cylindrical NaI(Tl) scintillation detector was placed next to the HPGe detector just outside the neutron beam as depicted in Figure~\ref{fig:setup}. Utilizing this detector, we performed a gamma coincidence analysis to measure the lifetime of the gamma cascade and extract the time delay between the nuclear recoil and the emission of the 68.75~keV gamma ray.

    \subsection{Digital Electronics}

    Both prior experiments~\cite{Jones1975,GeQF2021} utilized traditional analog shaping amplifiers and multichannel analyzers, whereby only shaped pulse amplitude was recorded. In this experiment, the raw detector output from both the HPGe and NaI(Tl) detectors were saved. Access to the HPGe preamplifier waveforms allowed a novel multi-shaping analysis by which we further studied the degree to which signal processing can bias the result.  
  
    Detector outputs were digitized at a rate of 100~MHz by a CAEN DT5780 module~\cite{CAEN1}. High-voltage bias and preamplifier power to the HPGe detector were provided by the same DT5780 unit. High voltage for the NaI(Tl) detector was provided by a separate CAEN DT5533E high-voltage module~\cite{CAEN2}. 
    %

    \subsection{Data Analysis}

    Raw preamplifier output waveforms were saved in 80~\textmu s-long traces. This allowed the same data to be processed and analyzed with multiple algorithms. Pulse amplitude was determined using three digital pulse shaping algorithms: the optimal filter~\cite{NumRecip,DSP,OF2021,NASAOF}, the trapezoidal filter, and a digitally synthesized CR-RC$^{8}$ Gaussian filter.
    
    The optimal filter was based on Refs.~\cite{NASAOF,OF2021} and constructed from a pulse template generated from averaging waveforms from the $^{57}$Co 122~keV calibration peak and noise waveforms acquired by random triggers. 
    
    The combined energy deposition by gamma ray and nuclear recoil was extracted from the optimally filtered data, while the other shaping filters facilitated quantification of the effect of analysis methodology on the result and comparison to prior studies.

    Gaussian shaping filters are commonly implemented by analog shaping amplifiers~\cite{Knoll2010,OrtecShaping}. A Canberra (Mirion) 2022 NIM spectroscopy amplifier with a shaping time constant of 8~\textmu s was used in the UChicago experiment~\cite{GeQF2021}. The amplifier used in the Brookhaven experiment~\cite{Jones1975} was not reported though the peaking time of 4~\textmu s was stated~\footnote[4]{The peaking time of a Gaussian filter is typically 2--2.5 times longer than the shaping time constant~\cite{OrtecShaping,Knoll2010}.}~\cite{Jones1975}. As both prior studies utilized multichannel analyzers, the maximum of the Gaussian-shaped signal within the saved trace window is taken as the pulse energy~\footnote[5]{The Brookhaven experiment does not explicitly report the use of a multichannel analyzer but it can be inferred as the x-axis of their energy spectra are labeled in the units of channel number.} The waveforms were processed with multiple shaping time constants to test the hypothesis that the choice of shaping time is the cause of the difference between the Chicago and Brookhaven results~\cite{GeQF2021,Jones1975}. 

    \begin{figure}[h]
        \includegraphics[width=.95\linewidth]{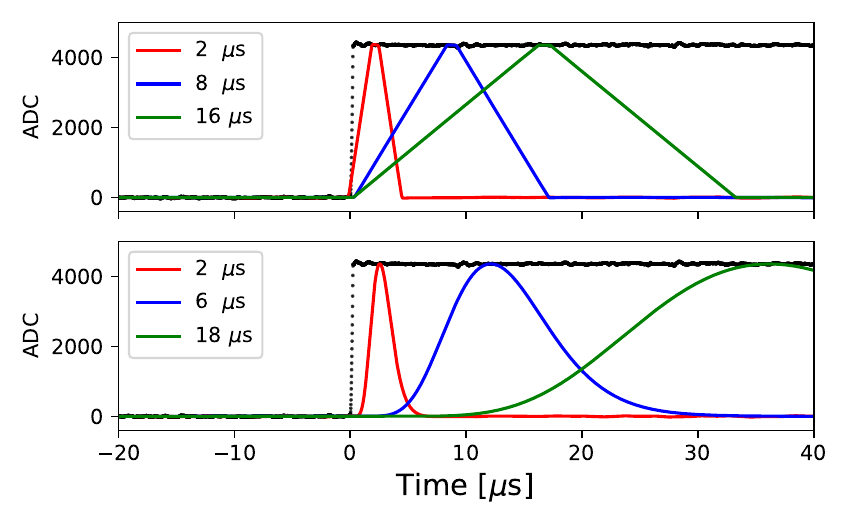}
        \caption{\label{fig:Shaping} Multiple shaping time trapezoidal (\textit{top}) and Gaussian / CR-RC$^{8}$ (\textit{bottom}) filters applied to the same digitized waveform from the HPGe detector (black). For a given shaping time, the peaking time is longer for the Gaussian filter than for the trapezoidal filter. This is because, for the Gaussian filter, the shaping time corresponds to the RC and CR time constants while for the trapezoidal filter, it corresponds to a linear integrating window width.}
    \end{figure}

    The trapezoidal filter was also implemented as it is typically the filter of choice for HPGe signal analysis~\cite{Knoll2010,IEEE,Keyser2013OptimizationOP}. The filter is well characterized and has been digitally implemented since the mid-1990s~\cite{JORDANOV1994261,JORDANOV1994337,Knoll2010,IEEE}. Three trapezoids with peaking times, of 2, 8, and 16~\textmu s were chosen to similarly test the effect of shaping time. An 800~ns flat-top time was chosen as it is sufficiently longer than the detector $\sim$100~ns (10--90\%) rise time, such that ballistic deficit is minimized~\cite{Knoll2010}. The amplitude of trapezoidally shaped signal was determined by sampling the middle of the flat top. The sample position was chosen by adding the associated peaking time and half the flat-top time to the pulse onset position. Sampling the flat-top is more robust to noise fluctuations than selecting the shaped signal maximum. 

    Minimal cuts were applied to the dataset in an attempt to not bias the result. Events were cut from the analysis based on saturation and pile-up. The effects of cuts within the energy range of interest are shown in Figure~\ref{fig:Spec1}. The pile-up cut minimally affects the region around the 68.752~keV gamma ray. The region below 67~keV is significantly affected due to the 66.725~keV state, which decays by emission of 53.4~keV and 13.3~keV gamma rays, the latter having a half-life of 2.91~\textmu s~\cite{PhysRev.174.331,PhysRevB.12.4793, PhysRevLett.32.512}.


    \subsection{Calibration}

    The energy scale of the HPGe detector was established using the 59.5409~keV gamma ray from $^{241}$Am and the Pb K$\alpha_{1}$ and K$\alpha_{2}$ X rays, 74.9694~keV and 72.8042~keV, respectively. A $^{57}$Co source and lead foil were used to produce X rays. The calibration sources were placed on top of a copper shim just above the beryllium window of the detector. The sources were present throughout the experiment to perform \textit{in-situ} calibration and monitor for gain drift. No such drift was observed during the experimental runs, but the calibration changed by $\sim$1\% from day to day as the detector was unbiased and re-biased over multiple days of measurements. 

    For the measurement of the lifetime of the nuclear states, the timescale between the HPGe and NaI(Tl) was calibrated utilizing a $^{22}$Na source. The energy scale of the NaI(Tl) detector was established with a $^{60}$Co gamma-ray source. Continuous calibration was not performed with the NaI(Tl) detector as it would have caused an increased rate of spurious coincidences. Mid-day calibrations delineated the experiment into seven approximately equal-sized data sets. The runs were analyzed separately and the results averaged.

\section{Results}

    \subsection{Energy Spectrum}
    
    The best resolution was achieved with the optimally filtered data set. We used this dataset for primary analysis, while the trapezoidally filtered and Gaussian-shaped datasets were used for assessing the systematic contribution from the choice of pulse processing methodology.
    
\begin{figure}[h]
        \includegraphics[width=.95\linewidth]{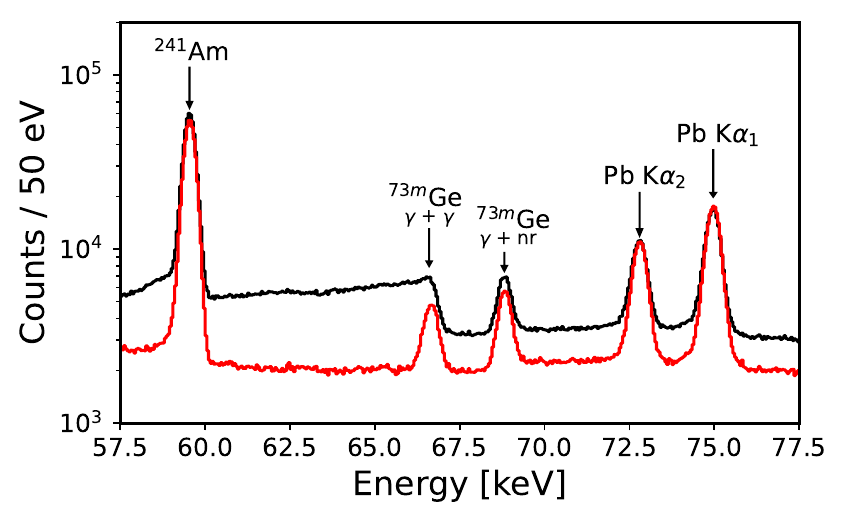}
        \caption{\label{fig:Spec1} Experimental energy spectrum in the energy range of interest with calibration and $^{72}$Ge (n,$\gamma$) peaks labeled as acquired (\textit{black}) and after pile-up cuts (\textit{red}).}
    \end{figure}

    In each of the data runs, the $^{241}$Am gamma peak and the Pb K$\alpha_{1}$ and K$\alpha_{2}$ X-ray peaks were fit with Gaussians modified by small correction functions based on Ref.~\cite{PeakFitting2}. The peak centroids were fit with a linear calibration function. These three peaks were chosen to provide the most accurate energy scale in the range of combined energy deposition by the 68.75~keV gamma ray and nuclear recoil and can be seen in Figure~\ref{fig:Spec1}. 
    The robustness of the energy scale was tested by performing an additional calibration using the 122~keV and 136~keV gamma-ray peaks. The change in scale was negligible compared to other uncertainties.
    The only feature strongly affected by the cuts is the aforementioned decay of the 66.7~keV state, which is not associated with the decay path of interest. The 66.7~keV peak was fit with an exponentially modified Gaussian functional form, the extracted energy from which was 66.720$\pm$0.011~keV, which agrees with the literature-reported value of 66.725$\pm$0.009~keV~\cite{NDS}.

    \subsection{Null Measurement}
    
    Thermal neutrons emitted from the beam port were verified to be the source of the signal. This was achieved by replacing the $^{6}$Li glass collimator with a solid $^{6}$Li glass glass disk for a two-hour data run.

\begin{figure}[h]
        \includegraphics[width=.95\linewidth]{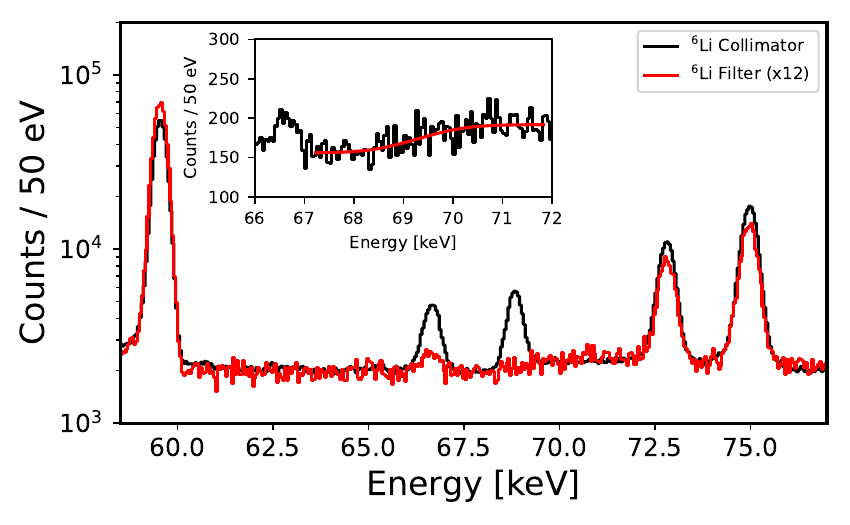}
        \caption{\label{fig:Null} Comparison of the null measurement (\textit{red}) to the thermal neutron data (\textit{black}). The null data set is scaled by a factor of 12 corresponding to the difference in experimental lifetimes. \textit{Inset}: step function fit to the fast neutron inelastic scattering structure (data un-scaled).}
    \end{figure}

    The null measurement is compared to the thermal neutron data in Figure~\ref{fig:Null}. The peak resulting from combined energy deposition of 68.75~keV gamma and nuclear recoil was not observed. A step-like feature was observed, which we hypothesize to be caused by the inelastic scattering of the fast neutron beam contribution able to pass through the $^{6}$Li glass filter. A similar feature is observed in the $^{6}$LiF shielded measurement reported in Figure~7 of Ref.~\cite{GeQF2021} and has been historically observed in other neutron experiments~\cite{SKORO1992333,JOVANCEVIC2010303}.

    A significantly reduced peak was observed at 66.7~keV associated with the corresponding level of $^{73m}$Ge. The kinetic energy imparted from fast neutron capture is not measured summed with the gamma-ray peak given the long, (0.5~s) lifetime of the 66.7~keV state. 
    

    \subsection{Fit Functional Form}

    The peak attributed to the combined energy deposition of a 68.75~keV gamma ray and nuclear recoil is fit with a Gaussian with a step to high energy:
    \begin{equation}
        f(x,\Vec{\mathbf{\pi}}) = \frac{A}{\sqrt{2\pi}\sigma}e^{-(x-\mu)^{2}/(2\sigma^{2})} + \frac{B}{2}\cdot\textbf{erfc}\bigg[ \frac{\mu - x}{\sqrt{2}\sigma}  \bigg] + C
        \label{eq:FitFunc}
    \end{equation}
    Parameters \textit{A} and \textit{B} are the area of the Gaussian and the amplitude of the step, respectively, while $\mu$ and $\sigma$ are the mean position and resolution; \textit{C} is the baseline offset term. The step function is motivated by the higher energy contribution of the neutron beam discussed above. The step position and smearing of the rising edge are fixed to the same centroid and resolution as the peak as justified in~\cite{PHILLIPS1976525,PeakFitting2}. 

\begin{figure}[h]
    \includegraphics[width=.95\linewidth]{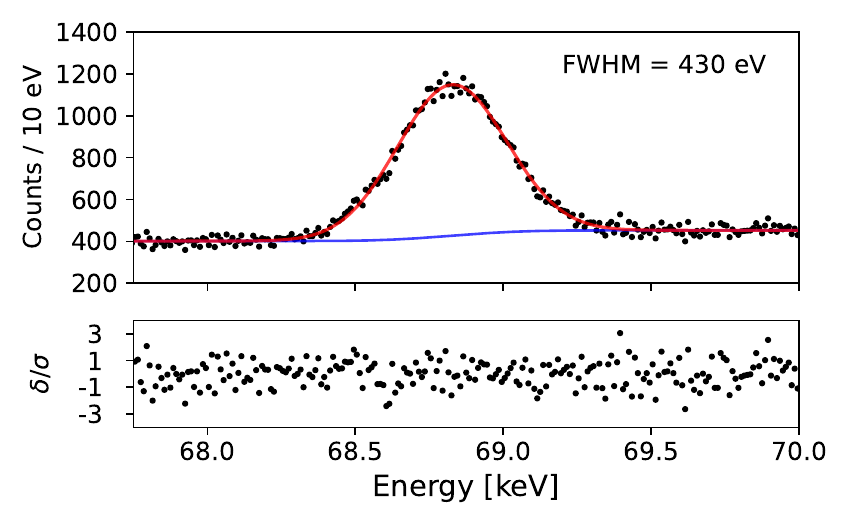}
    \caption{\label{fig:PeakFit} Fit (\textit{top}) and residual (\textit{bottom}) to the peak associated with the 68.75~keV gamma ray and nuclear recoil signal for all data sets. The red line shows the summed fit, Gaussian, baseline, and step while the blue trend is of the baseline and step components.}
    \end{figure}

    This functional form was fit to the entire data set as depicted with the residual in Figure~\ref{fig:PeakFit}. The fit was performed over 225 data points with an energy range of 67.75 to 70.0~keV. The $\chi^{2}$ value is 211.43 with a reduced $\chi^{2}$/NDF value of $211.43/(225-5)=0.96$, corresponding to a $\chi^{2}$-distribution value of 0.46. Given the goodness of the fit, additional terms were not added to the functional form so as to not over-fit the data. 

    \subsection{Ionization Energy and Uncertainty}
    
    The gamma plus nuclear recoil peak in each of the data sets was fit with the functional form described in Eq.~(\ref{eq:FitFunc}). The fit centroids, fit uncertainties, and calibration uncertainties are reported in Table~\ref{tab:means}. 
    
    The fit uncertainty, column~3, is the error on the fit centroid ($\mu$). The calibration uncertainty, column~4, is comprised of the errors in the slope and intercept parameters of the linear fit energy calibration. To conservatively account for the correlation between the slope and intercept terms, we sum the respective uncertainties. The combined error, column~5, of fit centroid plus calibration is calculated by summing the fit and calibration uncertainties in quadrature. 

    The fit values were averaged using an inverse error-squared weighted arithmetic mean yielding energy of 68.8158$\pm$0.0028~keV for the gamma ray plus ionization from the nuclear recoil.

    \begin{table}[h]
        \centering
        \begin{tabular}{|c|c|c|c|c|}
        \hline
        ~~Run~~  & ~~~E\textsubscript{$\gamma$} + E\textsubscript{nr}~~~        & Fit Unc. &  Calib. Unc. & Comb. Unc. \\ 
                              &  [keV]    & [eV]       &  [eV]  & [eV]        \\
        \hline
        \hline
        ~Run-1 & 68.8190 &   6.6 & 4.9 &  8.2  \\
        ~Run-2 & 68.8486 &  10.8 & 9.6 & 14.4   \\
        ~Run-3 & 68.8205 &   5.5 & 6.4 &  8.5  \\
        ~Run-4 & 68.8206 &   4.2 & 5.5 &  7.0 \\
        ~Run-5 & 68.8269 &   3.8 & 4.7 &  6.1 \\
        ~Run-6 & 68.7998 &   4.5 & 6.7 &  8.1  \\
        ~Run-7 & 68.8032 &   4.2 & 3.4 &  5.4 \\[0.1ex]
        \hline\hline
        ~Comb. & 68.8158 &   1.9  & 2.0 & 2.8 \\
        \hline
        \end{tabular}
        \caption{Mean parameter of the Gaussian fit to the combined energy deposition from 68.75~keV gamma ray and nuclear recoil with uncertainty and uncertainty from the calibration evaluated at the centroid position.}
        \label{tab:means}
    \end{table}

    %

    We estimate additional uncertainty contributions from the pulse selection cuts, choice of calibration peaks, and spread in values among the seven data sets. 
    
    The uncertainty from the pulse selection cuts is estimated by re-fitting the data without any cuts applied. The peak mean was found to be 68.8139~keV, a difference between this and our nominal value, 1.9~eV is taken as the uncertainty of the cuts. Similarly, a new calibration was performed including the 14.4 and 122~keV $^{57}$Co gamma rays. Under this new calibration, the peak is found to be at an energy of 68.8181~keV. We again take the difference, 2.3~eV, to be the uncertainty from the choice of calibration peaks. 
    
    The spread in extracted means among the seven data sets, 48.8~eV, is larger than the difference between the UChicago and Brookhaven results, 18.3~eV. To account for any possible unknown systematic effect causing the large spread, we re-calculate the error-weighted mean with the upper extreme value removed (run-2), the lower extreme value removed (run-6), and with both these runs removed. This procedure yielded new averages of 68.8145, 68.8179, and 68.8166~keV, respectively. The largest difference from outlier removal and our nominal energy of 68.8158~keV is 2.1~eV, which we take as a nuisance uncertainty.

    The uncertainty budget is reported in Table~\ref{tab:ub}. The budget includes uncertainty contributions (left-to-right): the fit centroid, calibration, pulse selection cuts, choice of calibration peaks, and the aforementioned nuisance uncertainty. The terms are summed in quadrature which yields a nominal energy value and uncertainty of 68.8158$\pm$0.0046~keV. 

    Subtracting the gamma energy of 68.752$\pm$0.007~keV yields a nuclear recoil ionization of 63.8$\pm$8.4~eV\textsubscript{ee}, corresponding to a quenching factor of 25.1$\pm$3.3\%.

    \begin{table}[h]
        \centering    
        \begin{tabular}{|c|c|c|c|c||c|}
        \hline
        Fit        & Calibration       & Pulse      & Calibration  & Nuisance & Total \\
        Centroid   & Fit               & Selection  & Selection    & Term     &       \\
        \hline
        1.9        & 2.0               & 1.9        & 2.3          & 2.1      & 4.6 \\
        \hline
        \end{tabular}
        \caption{Uncertainty budget for summed energy of 68.75~keV gamma ray with ionization produced by 254~eV\textsubscript{nr} nuclear recoil. Units are eV.}
        \label{tab:ub}
        
    \end{table}


    %

    


\section{Systematic Studies}

    \subsection{Lifetime of Nuclear States}
    
    \begin{figure}[h]
    \includegraphics[width=.95\linewidth]{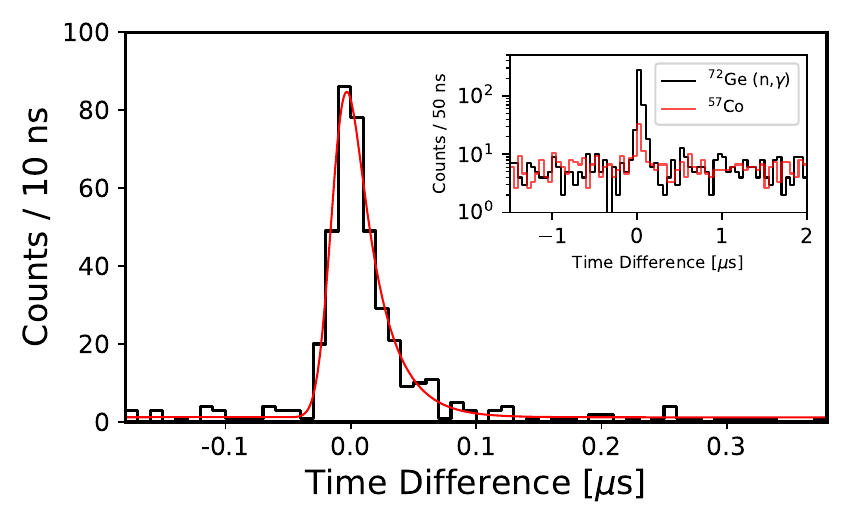}
    \caption{Time difference distribution between HPGe and NaI(Tl) detectors (\textit{black}) and fit (\textit{red}). Pulse selection cuts are applied as discussed in the text. \textit{Inset}: time difference distributions between detectors with energy gating around the 68.75~keV gamma plus recoil peak (\textit{black}) and 122~keV $^{57}$Co gamma peak to assess the spurious and accidental coincidences.\label{fig:TimePlot}}
    \end{figure}
    
    The Brookhaven experiment reported the lifetime of the 68.75~keV state to be 700~ns, and the preceding states in the de-excitation cascade have sub-nanosecond lifetimes~\cite{Jones1975}. We define the \textit{effective lifetime} of the 68.75~keV state as the sum of its lifetime and the lifetimes of the preceding 353~keV and 915~keV states (or 499~keV and 932~keV states, depending on the de-excitation path). This determines the delay between the nuclear recoil and the 68.753~keV gamma-ray signal within the HPGe detector. The UChicago study stipulates this 700~ns lifetime paired with the difference in time constants utilized in their shaping amplifiers to be the root cause for the difference between their and the Brookhaven result~\cite{GeQF2021}. 
    
    The effective lifetime is measured by finding time-coincident pairs of events between the HPGe and NaI(Tl) detectors. Energy gating is enforced to accept only coincident events where between 68.4~keV and 69.3~keV was deposited in the HPGe detector and more than 4~MeV was deposited in the NaI(Tl) detector. 
    
    A strong time-correlated signal is observed for the energy-gated events as shown in Figure~\ref{fig:TimePlot}; 373 coincident pairs were found between $-100$~ns and 200~ns. We define two background populations -- spurious and accidental coincident events. Spurious events are defined as randomly coincident events in HPGe and NaI(Tl) detectors originating from unrelated sources. This flat background is estimated to be 6~coincidence pairs per 100~ns. We define the accidental background as true coincident interactions between the two detectors, such as Compton scattering of external gamma rays that are not associated with the 68.75~keV emission from $^{73}$Ge.
    
    The accidental coincidence rate was estimated by applying a similar energy gate, 121.6~keV to 122.6~keV, around the 122~keV $^{57}$Co peak; 36 accidental coincident pairs were found after scaling for the peak area. Therefore, of the 373 time-tagged events, we estimate 319$\pm$30~counts originate from the $^{72}$Ge (n,$\gamma$) reaction, resulting in a signal-to-background ratio of $\sim$7:1. 
    
    A 100~MHz digitizer was selected for the experiment based on the expectation of a $\sim$700~ns state lifetime and was therefore sub-optimal for precisely measuring lifetimes shorter than several tens of ns. However, it can still be concluded that the lifetime is significantly shorter than 700~ns. The nuclear lifetime was evaluated by fitting the time difference distribution of coincident pairs to the function described in Eq.~\ref{eq:TimeFunc}.
    \begin{equation}
        f(x,\Vec{\mathbf{\pi}}) =  A\cdot e^{-(x-\mu)/\tau} \cdot\textbf{erfc}\bigg[ \frac{\mu - x}{\sqrt{2}\sigma}  \bigg] + C
        \label{eq:TimeFunc}
    \end{equation}
    The nominal fit values found a FWHM of 24~ns with a tail time constant of 22~ns. After de-convolution of the inherent detector response, determined from the $^{22}$Na calibration data, the effective lifetime was found to be $17.2\pm 7.6$~ns. This effective lifetime is $\sim$100~times shorter than the shaping time constant utilized in the Brookhaven experiment and would have a negligible effect on their result. The shaping time argument is, therefore, insufficient to explain the difference between the UChicago and Brookhaven results. 

    \subsection{Multi-Shaping Analysis}

    To further study the hypothesis of the impact of analysis methodology on the reconstructed combined energy deposition by the 68.75~keV gamma ray and nuclear recoil, waveforms were filtered by trapezoidal and Gaussian filters with varying time constants. The data was analyzed with three trapezoidal shaping times, 2, 8, and 16~\textmu s, and four Gaussian shaping times, 2, 6, 12, and 18~\textmu s.

    For both the trapezoidal and Gaussian shaping filters and all time constants, the reconstructed combined energy deposition by the 68.75~keV gamma ray and nuclear recoil differed by less than 2~eV with one exception, the 12~\textmu s Gaussian filter, where the difference is still within 2$\sigma$ of the accepted value. 
%

    \begin{figure}[h]
    \includegraphics[width=.95\linewidth]{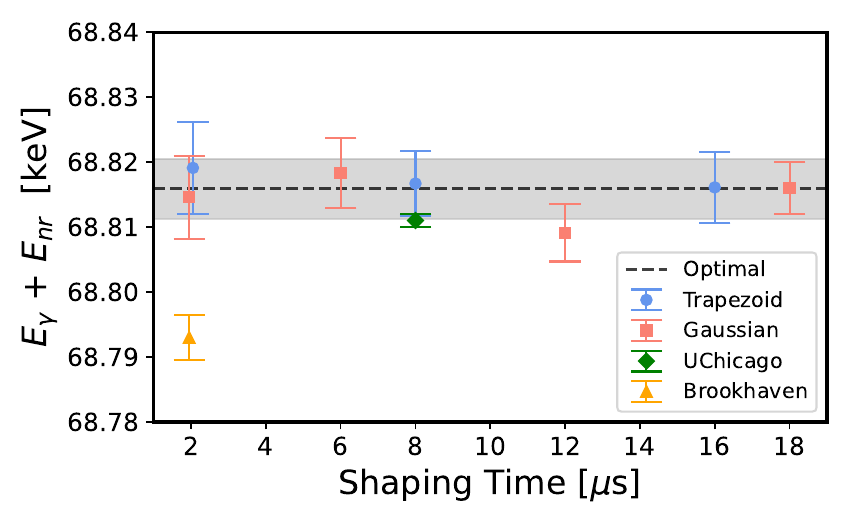}
    \caption{\label{fig:MultiShape} Combined energy deposition by the gamma ray and nuclear recoil evaluated by optimal filter (\textit{black}), trapezoidal filter (\textit{red}), and Gaussian filter (\textit{blue}) compared against the Brookhaven~\cite{Jones1975} and UChicago~\cite{GeQF2021} results. Optimal filtered uncertainty bounds (shaded grey region) include total uncertainty shown in Table~\ref{tab:ub}, while trapezoidal and Gaussian data points only include uncertainty contributions from calibration and peak fit.}
    \end{figure}
    
    The results of the multi-shaping analysis are shown in Figure~\ref{fig:MultiShape} with our accepted value from the optimal filter in black, trapezoidally filtered in blue, and Gaussian filtered in red. These results are compared against the Brookhaven~\cite{Jones1975} and UChicago~\cite{GeQF2021} results. The multi-shaping analysis does not explicitly account for statistical nor other systematic effects, but it demonstrates that the investigated analysis methodology, shaping filters, and shaping time constants do not explain the difference between the more recent studies (this work and UChicago experiment~\cite{GeQF2021}) and the earlier Brookhaven study~\cite{Jones1975}. This is further consistent with our measured lifetime of the relevant nuclear states, which decay on a time scale more $\sim$three orders of magnitude faster than the shaping time constants utilized in this study as well as in Refs.~\cite{Jones1975,GeQF2021}.

    \subsection{Position Dependence}
    
    One potential systematic effect discussed in neither of the two previous studies is the possibility of a calibration offset caused by the difference in the average location of $^{72}$Ge(n,$\gamma$) events compared to the calibration sources. The combined 68.75~keV gamma and nuclear recoil signal(s) are produced uniformly in the HPGe crystal whereas the $^{241}$Am gamma rays and Pb X rays predominantly interact near its top surface. We demonstrate no evidence for such position dependence through measurement of the decay of $^{71}$Ge produced in the crystal by neutron activation.
    
    \begin{figure}[h]
    \includegraphics[width=.95\linewidth]{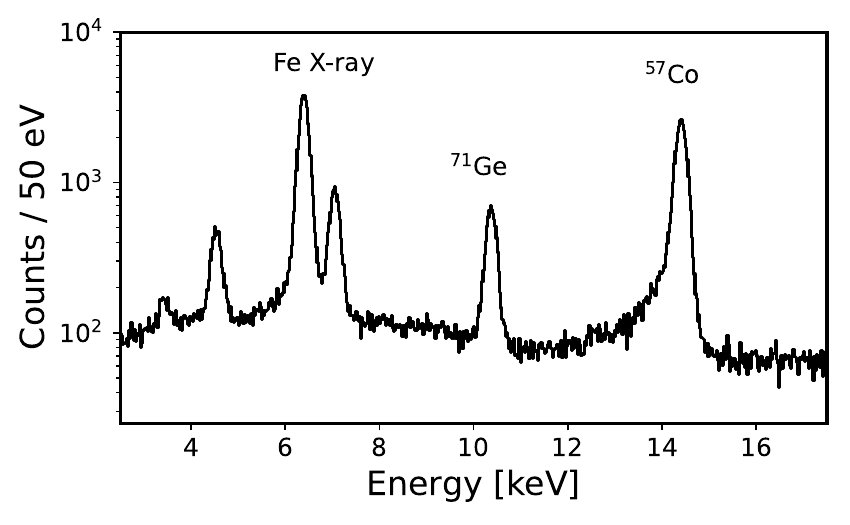}
    \caption{\label{fig:LowE} Low-energy spectrum following neutron irradiation. The $^{71}$Ge 10.37~keV peak is between the Fe X-ray peaks and the 14.4~keV gamma-ray peak, both from the $^{57}$Co calibration source.}
    \end{figure}

    Radioactive $^{71}$Ge decays via electron capture and emits 10.367~keV X rays corresponding to the K-edge of the daughter $^{71}$Ga~\cite{XRay,BEARDEN1967,FUGGLE1980275}. Similarly to the combined energy deposition by the $^{73m}$Ge gamma ray and nuclear recoil, the $^{71}$Ge decays are uniformly distributed throughout the depth of the HPGe crystal. 

    Following the neutron irradiation, we measured the 10.37~keV X rays, utilizing the 14.4~keV gamma ray, and 6.4~and~7.1~keV Fe X rays from $^{57}$Co for calibration. The $^{71}$Ge peak and calibration peaks analyzed with the same optimal filter analysis are shown in Figure~\ref{fig:LowE}. The $^{71}$Ge was fit with a Gaussian and a linear background term. The peak centroid was found to be 10.370~keV with a combined (fit and calibration) uncertainty of 4.4~eV. This value is within the uncertainty of both the accepted literature value of the $^{71}$Ga K-edge and the other measurements of $^{71}$Ge decay~\cite{XRay,BEARDEN1967,FUGGLE1980275,CoGeNT2011,Kang_2013}. Were there a calibration offset between events near the surface and in the bulk of the crystal, the effect would be even more evident for 10.37~keV than at 68.75~keV. From this analysis, we conclude that no significant systematic offset exists due to the event position dependence.

\section{Discussion and Conclusion}

    \subsection{Quenching Factor}
    
    The high level of agreement between our analysis methods shown in Figure~\ref{fig:MultiShape} gives confidence in the measured energy of the combined deposition by the 68.75~keV gamma ray and nuclear recoil to be 68.8158~$\pm$~0.0046~keV. The ionization produced by the nuclear recoil is calculated as the difference between this value and the 68.752~keV gamma-ray (alone) energy and is found to be: 63.8$\pm$8.4~eV\textsubscript{ee}. The quenching factor is then calculated as the ratio of this difference and 254.1~eV\textsubscript{nr} and yield 25.1~$\pm$~3.3\%. The accurate value for the energy of the gamma ray is therefore critical for determining the quenching factor, as demonstrated in Table~\ref{tab:results}. 

    Our results are consistent within the 90\% confidence interval recently published in Ref.~\cite{Bonhomme2022}. The UChicago photonuclear data, which suggests a deviation from the Lindhard model $\lesssim$1.5~keV\textsubscript{nr}, is similarly consistent within the confidence interval of Ref.~\cite{GeQF2021,Bonhomme2022}. However, the UChicago iron filter (Fef) data suggest a quenching factor greater than 30\% if projected down to 254~eV\textsubscript{nr}.
    %
    
    \begin{table}[h]
        \centering
        \begin{tabular}{|c|c|c|c|c|}
        \hline
        Study  & $\gamma$ + Recoil & $\gamma$ &  Ionization & Quenching  \\ 
               & [keV]   &   [keV]  & [eV\textsubscript{ee}]    &       [\%]          \\
        \hline
        \hline
        ~This Work  &~68.816~&~68.752~& $64\pm8$ & $25.1\pm3.3$ \\
        \hline
        ~UChicago   &~68.811~&~68.752~& $59\pm7$  & $23.2\pm2.8$ \\
        \hline
        ~Brookhaven &~68.793~&~68.752~& $41\pm8$ & $16.0\pm3.1$  \\
        \hline
        ~Lindhard   & ~-~    & ~-~    & 36.2     & 14.3        \\
        \hline
        \multicolumn{5}{@{}l}{\makecell{As appears in original work:}}\\
        \hline
        ~UChicago   &~68.811~&~68.734~& $77\pm20$  & $30.3\pm8$ \\
        \hline
        ~Brookhaven &~68.793~&~68.754~& $39\pm6$ & $15.6\pm2$  \\
        \hline
        \end{tabular}
        \caption{Comparison of ionization yields and quenching factors calculated with the ENSDF gamma-ray energy of 68.752~keV. The lower portion of the table lists quenching data as originally reported in Refs.~\cite{Jones1975} and~\cite{GeQF2021}.}
        \label{tab:results}
    \end{table} 
        
    \subsection{Interatomic Potential}
    
    The agreement between experimental data and the Lindhard model, $\kappa=0.157$, observed $\gtrsim$10~keV\textsubscript{nr} but poor agreement observed near $\sim$1~keV\textsubscript{nr}~\cite{Jones1965,Jones1967,Jones1968,Jones1971,Jones1975,MESSOUS1995,SIMON2003643,Barbeau_2007,EDELWEISS2007,CDMS2010QF,SOMA201667,Scholz2016,GeQF2021,Bonhomme2022} suggest a significant modification in the quenching physics near the keV\textsubscript{nr} energy scale in Ge, which may be related to its material (crystalline) structure. 
    
    Both this work and the UChicago study~\cite{GeQF2021} indicate a significant enhancement in the quenching factor in Ge above the Lindhard model. Ref.~\cite{GeQF2021} postulates the Migdal effect as a potential explanation. However, more recent measurements find no Migdal nor Migdal-like signal in xenon~\cite{JingkeXenon2023}; as a result, any modification to the Lindhard theory requires a material-dependent model. Ref.~\cite{PhysRevD.106.043004} discusses critical differences between isolated atoms and those bound to a lattice within the context of the Migdal effect. 
    
    We hypothesize the interatomic potential plays a key role in the ionization produced by low-energy, $\lesssim$1--2~keV\textsubscript{nr} nuclear recoils. The stopping distance for a 254~eV\textsubscript{nr} $^{73}$Ge ion in Ge is of order 10~{\AA} as discussed in Ref.~\cite{Jones1975}. This is the same length scale as the Ge lattice spacing, and hence the fundamental assumptions of Lindhard theory~\cite{Lindhard}, notably that the ion travels through a homogeneous medium are not met. Furthermore, the vacancy site produced in the local lattice necessitates a change to the interatomic potential and band gap energy.
    
    \subsection{Conclusion}
    
    This work reports a measured ionization yield of $63.8 \pm 8.4$~eV\textsubscript{ee} from monoenergetic 254.1~eV\textsubscript{nr} Ge nuclear recoil. This corresponds to a quenching factor of $25.1\pm3.3$~\%, which confirms the 2021 UChicago study~\cite{GeQF2021}. We have expanded upon it by tagging events known to be correlated with the emission of the 5.8~MeV gamma ray and hence must contain the signal from the 254~eV\textsubscript{nr} nuclear recoil. We have also shown the effective lifetime between the nuclear recoil and the decay of the 68.75~keV state to be shorter than $\sim$700~ns estimated in Ref.~\cite{Jones1975} and discussed in Ref.~\cite{GeQF2021}. By saving raw preamplifier outputs, we demonstrated that shaping time and analysis methodology do not affect the result to the degree necessary to explain the discrepancy between the UChicago and Brookhaven studies. Lastly, by measuring the 10.37~keV signal from the decay of $^{71}$Ge, we have demonstrated no significant calibration offset originating from the interaction position in the HPGe crystal.
    
    Additional accurate measurement of the energy of the 68.75~keV nuclear level of $^{73}$Ge is motivated as it impacts the quenching factor value and is also the leading source of uncertainty. However, broad systematic offsets would be required in the nuclear data~\cite{NDS,IAEANDS} to reconcile this work and the UChicago study with the Lindhard model. 


\acknowledgments{This work was supported by the U.S. Department of Energy, Office of Nuclear Energy under DOE Idaho Operations Office Contract DE-AC07- 051D14517 as part of a Nuclear Science User Facilities experiment. This work was supported in part by the Department of Energy National Nuclear Security Administration, Consortium for Monitoring, Verification and Technology (DE-NE000863) and the Department of Energy National Nuclear Security Laboratory Research Graduate Fellowship. We thank the staff of the Ohio State University Nuclear Reactor Laboratory, Raymond Cao, Kevin Herminghuysen, Andrew Kauffman, Maria McGraw, Matthew Van Zile, and Susan White for their assistance in coordinating, setting up, and operating the experiment. We would also like to thank William Heroit of the University of Michigan for his assistance. Lastly, we thank thank Connor Bray and Jingke Xu for numerous productive conversations.}

\nocite{*}

\bibliography{Bib}

\end{document}